\documentclass[times,astrobib,amssymb,usenatbib]{mn2e}
\usepackage{graphicx}
\usepackage{txfonts}
\usepackage{natbib}
\usepackage{supertabular}
\usepackage{longtable}
\usepackage{url}
\usepackage{dcolumn}
\usepackage{placeins}
\usepackage{multirow} 
\usepackage{rotating}
\begin{document}
\newcommand{\sqcm}{cm$^{-2}$}  
\newcommand{\lya}{Ly$\alpha$}
\newcommand{\lyb}{Ly$\beta$}
\newcommand{\lyg}{Ly$\gamma$}
\newcommand{\heo}{\mbox{He\,{\sc i}}}
\newcommand{\he}{\mbox{He\,{\sc ii}}}
\newcommand{\hi}{\mbox{H\,{\sc i}}}
\newcommand{\hw}{\mbox{H\,{\sc ii}}}
\newcommand{\oth}{\mbox{O\,{\sc iii}}}
\newcommand{\ofo}{\mbox{O\,{\sc iv}}}
\newcommand{\of}{\mbox{O\,{\sc v}}}
\newcommand{\os}{\mbox{O\,{\sc vi}}}
\newcommand{\ose}{\mbox{O\,{\sc vii}}}
\newcommand{\oei}{\mbox{O\,{\sc viii}}}
\newcommand{\cf}{\mbox{C\,{\sc iv}}}
\newcommand{\cfi}{\mbox{C\,{\sc v}}}
\newcommand{\csi}{\mbox{C\,{\sc vi}}}
\newcommand{\cto}{\mbox{C\,{\sc ii}}}
\newcommand{\ct}{\mbox{C\,{\sc iii}}}
\newcommand{\sito}{\mbox{Si\,{\sc ii}}}
\newcommand{\sit}{\mbox{Si\,{\sc iii}}}
\newcommand{\sif}{\mbox{Si\,{\sc iv}}}
\newcommand{\nt}{\mbox{N\,{\sc iii}}}
\newcommand{\nfo}{\mbox{N\,{\sc iv}}}
\newcommand{\nf}{\mbox{N\,{\sc v}}}
\newcommand{\neo}{\mbox{Ne\,{\sc viii}}}
\newcommand{\nete}{\mbox{Ne\,{\sc x}}}
\newcommand{\mgt}{\mbox{Mg\,{\sc ii}}}
\newcommand{\fet}{\mbox{Fe\,{\sc ii}}}
\newcommand{\mgx}{\mbox{Mg\,{\sc x}}}
\newcommand{\zabs}{$z_{\rm abs}$}
\newcommand{\zmin}{$z_{\rm min}$}
\newcommand{\zmax}{$z_{\rm max}$}
\newcommand{\zqso}{$z_{\rm qso}$}
\newcommand{\subHe}{_{\it HeII}}
\newcommand{\subH}{_{\it HI}}
\newcommand{\subHLy}{_{\it H Ly}}
\newcommand{\degree}{\ensuremath{^\circ}}
\newcommand{\lapp}{\mbox{\raisebox{-0.3em}{$\stackrel{\textstyle <}{\sim}$}}}
\newcommand{\gapp}{\mbox{\raisebox{-0.3em}{$\stackrel{\textstyle >}{\sim}$}}}
\newcommand{\be}{\begin{equation}}
\newcommand{\en}{\end{equation}}
\newcommand{\di}{\displaystyle}
\def\tworule{\noalign{\medskip\hrule\smallskip\hrule\medskip}} 
\def\onerule{\noalign{\medskip\hrule\medskip}} 
\def\bl{\par\vskip 12pt\noindent}
\def\bll{\par\vskip 24pt\noindent}
\def\blll{\par\vskip 36pt\noindent}
\def\rot{\mathop{\rm rot}\nolimits}
\def\alf{$\alpha$}
\def\refff{\leftskip20pt\parindent-20pt\parskip4pt}
\def\kms{km~s$^{-1}$}
\def\zabs{$z_{\rm abs}$}
\def\zem{$z_{\rm em}$}
\title[Highly ionized outflow]{Highly ionized collimated outflow from HE~0238--1904}
\author[S. Muzahid et al.]
{
\parbox{\textwidth}{ 
S. Muzahid$^{1}$, 
R. Srianand$^{1}$, 
B. D. Savage$^{2}$, 
A. Narayanan$^{3}$, 
V. Mohan$^{1}$  
and G. C. Dewangan$^{1}$  
} 
\vspace*{4pt}\\  
$^{1}$ Inter-University Centre for Astronomy and Astrophysics, Post Bag 4, 
Ganeshkhind, Pune 411\,007, India \\ 
$^{2}$ Department of Astronomy, University of Wisconsin, 475 North Charter Street, 
Madison, WI, 53706, USA \\ 
$^{3}$ Indian Institute of Space Science \& Technology, Thiruvananthapuram 695547, 
Kerala, India
}
\date{Accepted. Received; in original form }
\maketitle
\label{firstpage}
\begin {abstract} 
We present a detailed analysis of a highly ionized, multiphased and collimated 
outflowing gas detected through \of, \os, \neo\ and \mgx\ absorption associated 
with the QSO HE~0238--1904 (\zem $\simeq 0.629$). 
Based on the similarities in the 
absorption line profiles and estimated covering fractions, we find that the \os\ and 
\neo\ absorption trace the same phase of the absorbing gas. Simple photoionization 
models can reproduce the observed $N(\neo)$, $N(\os)$ and $N(\mgx)$ from a single 
phase whereas the low ionization species (e.g. \nt, \nfo, \ofo) originate from a 
different phase. The measured $N(\neo)/N(\os)$ ratio is found to be remarkably 
similar (within a factor of $\sim$ 2) in several individual absorption components 
kinematically spread over $\sim 1800$~{\kms}. Under photoionization this requires a 
fine tuning between hydrogen density (n$_{\rm H}$) and the distance of the 
absorbing gas from the QSO. Alternatively this can also be explained by collisional 
ionization in hot gas with $T \ge 10^{5.7}$~K. Long-term stability favors the 
absorbing gas being located outside the broad line region (BLR). We speculate that 
the collimated flow of such a hot gas could possibly be triggered by the radio jet 
interaction. 
\end {abstract}
\section{Introduction} 
Large scale outflows from AGNs play a vital role in regulating star formation 
in galaxies and the growth of the super-massive black holes at their centers 
\citep[]{Silk98,King03,Bower06}. Hence detecting different forms of outflows is 
essential to understand the AGN feedback. In the QSO spectrum, outflows can 
manifest as associated narrow absorption lines (NALs) or broad absorption lines 
(BALs) in the ultraviolet (UV) and as high ionization absorption lines and edges 
(i.e. warm absorbers, WAs) in the soft 
X-ray wavelengths. BALQSOs comprise of $\sim$ 40\% of the total QSO population 
\citep[][]{Dai08} while WAs are detected in the X-ray spectrum of $\sim 50\%$ 
of Seyfert galaxies \citep{Crenshaw03} and QSOs \citep{Piconcelli05}. 
BALs \& NALs are predominantly detected through species with ionization 
potential (IP) $\lesssim 100$ eV (e.g. \cf, \sif, \nf\ etc.), WAs, on the 
contrary, are identified by species with IP $\gtrsim 0.5$ keV 
(e.g. \ose, \oei\ etc.). To understand the ionization structure of the 
outflowing gas entirely it is very important to detect species with 
intermediate ionization potentials such as \os\ (138 eV), \neo\ (239 eV), 
\mgx\ (367 eV) etc. In particular the resonant lines of \os $\lambda\lambda 1031,1037$ 
\neo $\lambda\lambda 770,780$ and \mgx $\lambda\lambda 609,624$ are well 
suited for probing the intermediate physical conditions of the outflowing gas. 
However the detections of such species in outflows have been very rare till date. 

There are three confirmed and two tentative detections of associated \neo\ 
(most of them showing radio emission) in the form of narrow absorption lines 
[UM~675, \citet{Hamann95}; 3C ~288.1, \citet{Hamann00}, J2233$-$606, 
\citet{Petitjean99}; HE 0226-4110, \citet{Ganguly06} and
3C48, \citet{Gupta05}] and three QSOs in the form of BAL absorption 
[Q~0226$-$1024, \citet{Korista92}; SBS~1542+541, \citet{Telfer98} 
and PG~0946+301, \citet{Arav99}]. 
While multiphase photoionization models are generally used to explain these 
observations \citep[see for example,][]{Hamann97}, the role of collisional 
ionization is not adequately explored. 
Here we report the detection of a very strong associated \neo\ (and \os) absorption 
in the high signal-to-noise ($S/N$) $HST$/COS and $FUSE$ spectra of HE 0238$-$1904 
which has spectral energy distribution (SED) typical of a non-BALQSO. 
The system is at an ejection velocity of $\sim$ 4500 \kms\ with a kinematic spread 
of $\sim$2500 \kms. Along with the \neo\ and \os, we also detect absorption from 
\nt, \nfo, \ofo, \of\ and \mgx. Throughout this paper we use cosmology with 
$\Omega_{\rm M}$ = 0.27, $\Omega_{\Lambda}$ = 0.73 and $H_0$ = 71 \kms Mpc$^{-1}$. 
\section{Observations of HE~0238$-$1904} 
\label{obs} 
We use $HST$/COS and $FUSE$ FUV and NUV spectroscopic observations of 
HE~0238$-$1904 that are publicly available in the MAST archive\footnote{http://archive.stsci.edu/}. The $FUSE$ observations were carried out during 
December 2000 (PI, Moos), July 2003 and October 2004 (PI, Howk). We have 
downloaded a total of 86 frames calibrated using 
CALFUSE (v3.2.1) pipeline. The combined spectrum has a resolution of 
$R \sim 10,000$ and S/N ratio of 3$-$10 per pixel over the wavelength range 920$-$1180 $\AA$ 
\citep[]{Moos00,Sahnow00}. The archived COS observations are from the COS 
GTO Program 
11541 (PI, Green). These observations consist of G130M and G160M 
FUV grating integrations at medium resolution of $R \sim 20,000$ and S/N $\sim$ 
20$-$30 per pixel in the wavelength range of 1134$-$1796$~{\AA}$ \citep[]{Osterman11,Green12}. 
These spectra and the low-dispersion ($R \sim 3000$) NUV spectra covering the wavelength 
range of 1650$-$3200$~{\AA}$ obtained with the G230L COS grating from 
the archive were extracted  using the CalCOS v2.15.4 pipeline. Continuum normalization 
was done by fitting the regions free of absorption lines with a smooth lower order 
polynomial. 

The optical spectrum of HE~0238$-$1904 was obtained with the 2-m telescope 
at IUCAA Girawali Observatory (IGO)\footnote{http://www.iucaa.ernet.in/$\sim$itp/} on 7-8 December 2011. The long slit spectra (3x30 
min exposures), covering the  wavelength range $3000-9000~\AA$, were obtained with 
the GR5 grism of the IUCAA Faint Object Spectrograph (IFOSC) using a slit width of 
1.5 arcsec. The raw CCD frames were cleaned and the spectra were extracted using 
standard IRAF \footnote{IRAF is distributed by National Optical Astronomy 
Observatories, which are operated by the Association of Universities for Research 
in Astronomy, Inc., under cooperative agreement with the National Science Foundation.} 
procedures. We use the method explained in \citet{Vivek09} for removing the fringing 
at $\lambda > 7000~\AA$.  The wavelength and flux calibrations were performed using 
Helium-Neon lamps and standard star spectrum respectively. The final co-added 
spectrum (in the heliocentric frame) has a resolution of $R \sim 300$ and a S/N 
$\sim 20-40$ per pixel. 
\subsection{Spectral Energy Distribution and QSO parameters} 

From simultaneous Gaussian fits to $\rm H\beta$, $\oth~\lambda\lambda~ 4960,5008$ emission 
lines we get the emission redshift, \zem\ = $0.629\pm0.002$.  
The rest-frame UV/optical Spectral Energy Distribution (SED) of 
HE~0238$-$1904 is shown in Fig.~\ref{sed}. The (green) star is the X-ray 
flux calculated from XMM-Newton slew survey data assuming 
$f_{\nu} \propto \nu^{-0.7}$ \citep{Sambruna99} between 0.2 to 2 keV and 
Galactic $N(\hi) = 3\times10^{20} \rm cm^{-2}$. 
Consistency of photometric points (purple triangles) of  \citet{Ojha09}  with our 
IGO spectrum (in red) implies lack of  
strong flux variability over 4.2 yrs in QSO rest frame.
This is also supported by the similar UV flux measurements made by 
COS in December 2009 and STIS G140L in July 2002. In the absence of 
strong variability, multiple epoch data can be combined to get the QSO's 
SED. Fig.~\ref{sed} shows that the source is 
intrinsically bringhter in UV than the HST composite spectrum of radio-loud QSOs 
constructed by \citep{Telfer02}.

We measure X-ray-to-optical spectral index, $\alpha_{ox} = -1.60$ which is consistent 
with $\alpha_{ox} = -1.55$ we get from the measured optical luminosity $L_{2500}$ 
using equation 8 of \citet{Stalin10}. This indicates that HE~0238$-$1904 is more like 
a typical non-BAL QSO. 
The full width at half maxima (FWHM) of the $\rm H\beta$ line is $\sim 7200$ \kms. 
Using the prescription of \citet{Bentz09} we measure the size of the BLR to be 
$R_{\rm BLR} \sim 1.1\times10^{18}$ cm. This together with the 
FWHM of $\rm H\beta$ gives the black hole mass 
M$_{\rm BH}$ = 2.4$\times10^{10}$~M$_{\odot}$ \citep[]{Onken04} and a Schwarzschild 
radius of $R_{\rm Sch} \sim 7.0\times10^{15}$ cm. Following \citet{Hall11}, we find 
the diameter of the disc within which 90\% of the 2700~$\AA$ continuum is emitted 
$D_{2700} \sim 3.2\times10^{17}$ cm. 
\section{Analysis of the \neo\ absorber} 
\label{ana} 
\begin{figure} 
\centerline{
\vbox{
\centerline{\hbox{ 
\includegraphics[height=9.cm,width=8.0cm,angle=270]{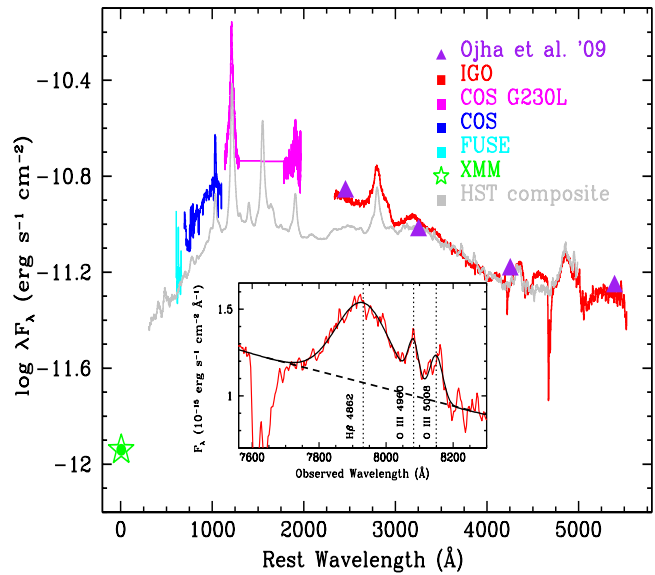} 
}}
}}
\caption{Rest-frame SED of HE~0238$-$1904. The green 
star is from XMM slew survey data. The spectrum in cyan, blue, magenta 
and red are obtained from FUSE, COS(G130M, G160M), COS(G230L), and IGO 
respectively. The purple triangles are the photometric points 
from \citet{Ojha09}. The HST composite spectrum for radio-loud QSOs is plotted 
in light gray with arbitrary normalization. Inset shows Gaussian fits to H$\beta$, 
and \oth\ $\lambda\lambda$4960,5008 lines which are used to estimate the
emission redshift.  
} 
\label{sed} 
\end{figure}

\begin{figure*} 
\centerline{
\vbox{
\centerline{\hbox{ 
\includegraphics[height=18.0cm,width=10.0cm,angle=-90]{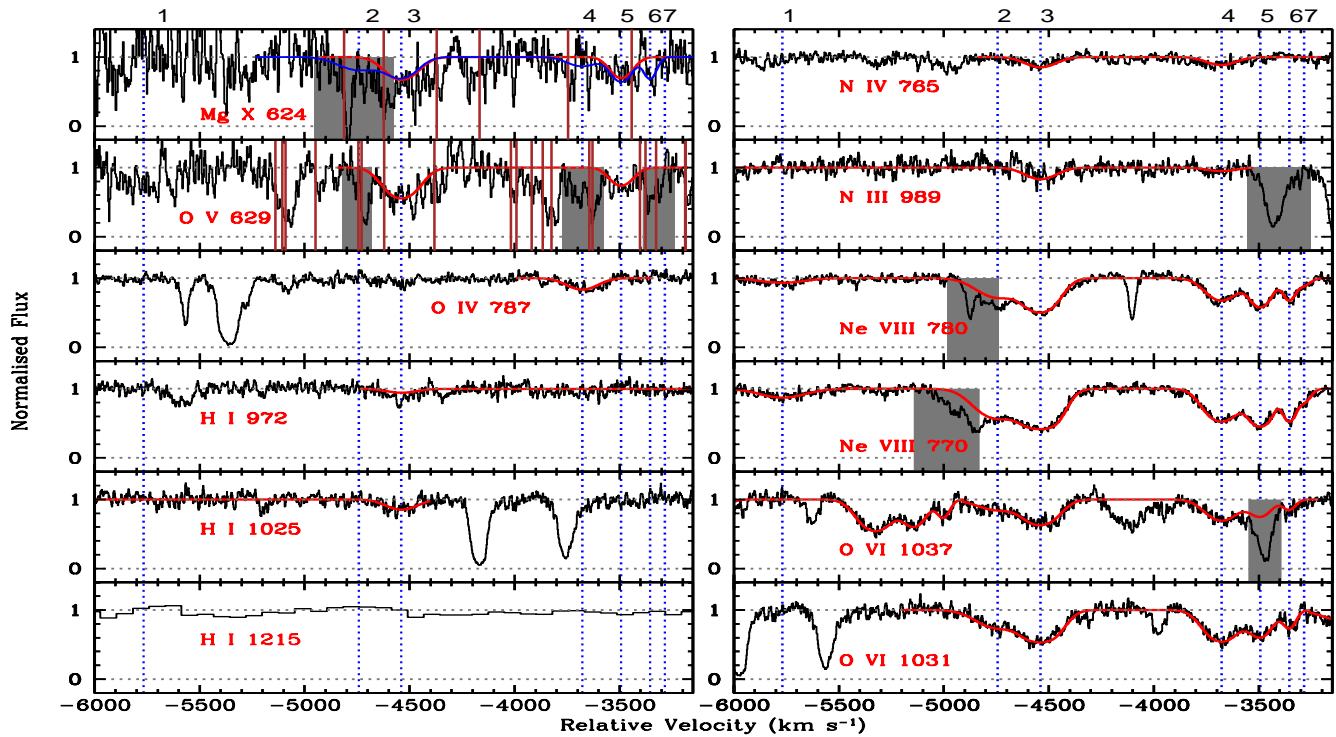} 
}}
}}
\caption{Absorption profiles of different species for the associated absorber 
shown with respect to the ejection velocity from the QSO. The \lya\ is from low 
dispersion COS (G230L) spectrum whereas \mgx\ and \of\ are from FUSE. In total 
there are seven \neo\ components identified by the vertical dotted lines. The best 
fitting Voigt profiles after correcting for partial coverage are overplotted as 
smooth red curves. Absorption lines unrelated to this system are marked by the 
shaded regions. The blue smooth curve in the \mgx\ panel is the synthetic profiles 
with column densities as predicted from the photoionization model (see text). 
The vertical solid lines in the \mgx\ and \of\ panels show the positions of 
Galactic H$_{2}$ lines. 
} 
\label{vplot} 
\end{figure*} 

\begin{figure} 
\centerline{
\vbox{
\centerline{\hbox{ 
\includegraphics[height=8.4cm,width=9.0cm,angle=00]{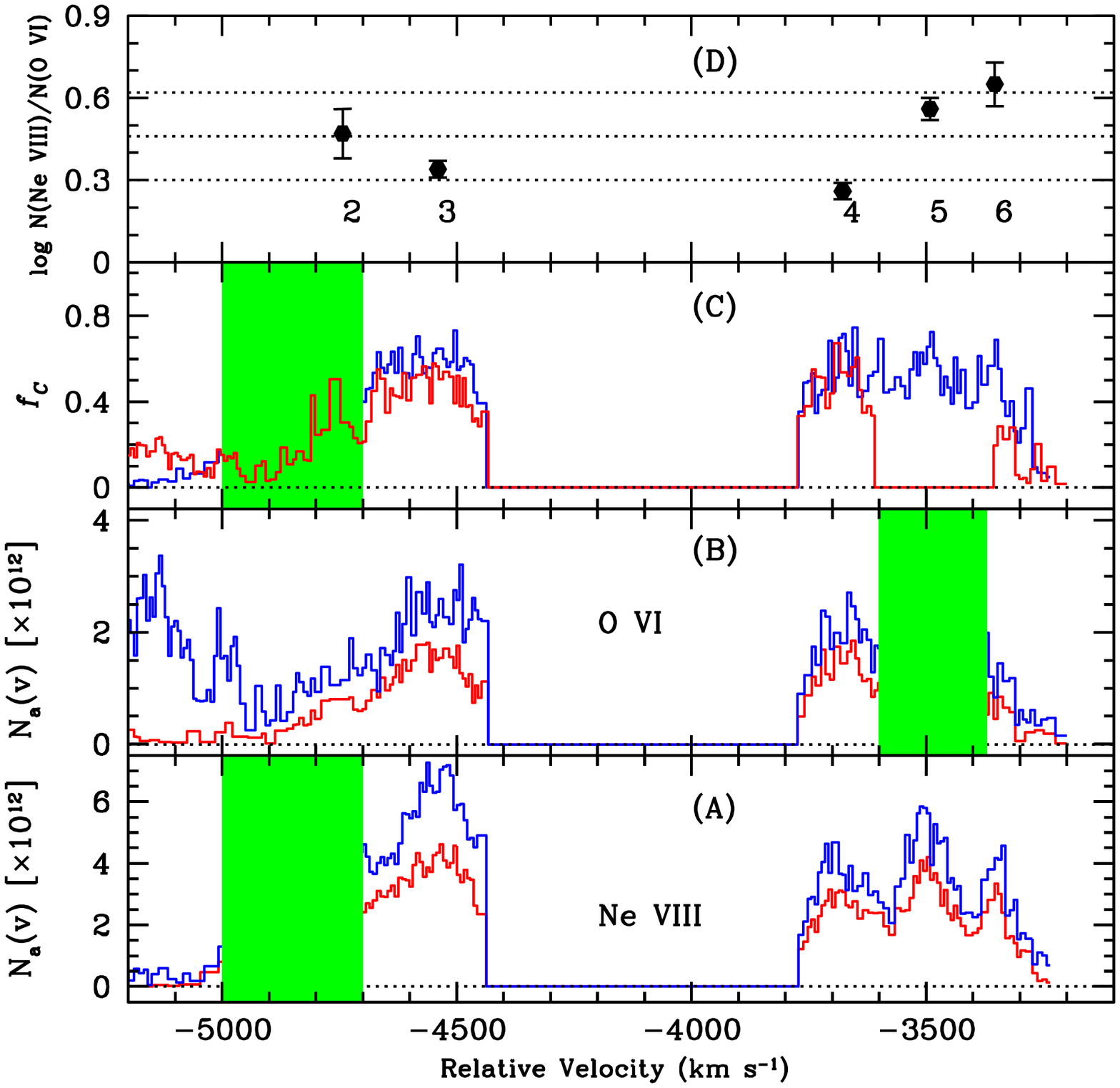} 
}}
}}
\caption{   Panel (A) and (B) : apparent column density profiles of \neo\ and \os\  
            doublets respectively. The stronger and weaker transitions of the 
            doublets are shown in red and blue respectively. 
	    Panel (C) : distributions of $f_c$ for \os\ (in red) 
	    and \neo\ (in blue). Panel (D) : $N(\neo)/N(\os)$ measured in individual 
	    components labeled by the numbers. The mean and 1$\sigma$ scatter of 
	    the ratio are shown in horizontal dotted lines. The contamination are 
	    shown by the shaded regions.   
	    } 
\label{cov_frac} 
\end{figure} 

In Fig.~\ref{vplot} we show absorption profiles of different species as a function 
of outflow velocity with respect to the QSO (\zem\ = 0.629). The \neo\ absorption is 
detected in seven components spread over a $\sim 2500$~{\kms} velocity interval. 
The separate components are labeled in the figure. Apart from the two weak \neo\ 
components (1 \& 7), all other components  show \os\ absorption with profiles very 
similar to that of \neo. The high ionization \mgx $\lambda 624$ and \of $\lambda 629$ 
lines  are clearly detected in the $FUSE$ data but without blending only in 
components 3 \& 5. 
Due to poor spectral S/N the presence of stronger \mgx $\lambda 609$ can be confirmed 
only for component 3. 
Contamination from Galactic H$_2$ 
lines has made the detection of \mgx\ and \of\ ambiguous in all other components. 
At the velocities of components 3 \& 4, we also find weak absorption consistent with 
\nt $\lambda 989$ and \nfo $\lambda 765$. The stronger \nt $\lambda 685$ line 
($f \lambda \sim 1.7$ times) is a non-detection in the $FUSE$ spectrum, and 
therefore we report the $N(\nt)$ measurements for these components as upper limits. 
Very weak Ly$\beta$ and Ly$\gamma$ absorption are seen in the COS data for 
component 3. But we do not find any Ly$\alpha$ absorption in the low dispersion 
COS G230L spectrum in any of these \neo\ components.

The apparent column density profiles of \neo\ and \os\ doublets,  shown in 
Fig.~\ref{cov_frac}, clearly indicates that the absorber is only partially covering 
the background emitting source. The estimated covering fractions ($f_{c}$) using the 
method described in \citet{Srianand99} are also shown in Fig.~\ref{cov_frac}. The 
covering fractions estimated for the \neo\ in components 2$-$7 is $\sim$0.6. The 
\os\ absorption shows $f_c \sim 0.5$ in components 2 \& 3 and $f_c \sim 0.6$ in 
components 4$-$6. The weak \neo\ absorption in component 1 is consistent with a 
complete coverage of the background source. It is interesting to note that \neo\ 
and \os\ ions have very similar covering fractions for all the components. This 
together with similar profile shapes suggest that the two ions are possibly tracing 
the same phase of the absorbing gas. The covering fraction of $f_c \sim 0.1$ for 
\hi\ determined using Ly$\beta$ \& Ly$\gamma$ of component 3 is significantly 
smaller implying the kinematic coincidence of multiple gas phases with different 
projected area \citep[see also][]{Telfer98}. The results of Voigt profile fitting 
analysis after taking into account these effects of partial coverage is given in 
Table~\ref{tab_cl}. In case of non-detections we calculate 3$\sigma$ upper limits on 
the column densities using the rms error in the unabsorbed continuum at the expected 
position and by using the $b$ value as measured for \neo\ (or \os) with the 
appropriate $f_c$. 

The most intriguing fact about this system is that in the five components where we 
have measurements of $N(\neo)$ and $N(\os)$, we find  a surprisingly constant value 
of 0.46$\pm$0.16 for the $\log~N(\neo)/N(\os)$ ratio (see panel-D of Fig.~\ref{cov_frac}). 
To understand the physical conditions of the absorbing gas we ran several 
photoionization models with Cloudy v(07.02) \citep[]{Ferland98}. In these models we 
assumed (a) the ionizing spectrum to be a combination of black body 
(with $T\sim1.5\times10^{5}$~K) and power law with $\alpha_{x} = -0.7$, 
$\alpha_{uv} = -0.5$ and $\alpha_{ox} = -1.6$ as observed for HE~0238$-$1904 
(see Fig.~\ref{sed}), (b) the gas is an optically thin plane parallel slab and 
(c) the relative abundances of heavy elements are similar to the solar values 
\citep[]{Asplund09}. 

In Fig.~\ref{model} various model predicted column density ratios are plotted 
against ionization parameter. In the case of components 3 \& 5, the observed 
$N(\neo)/N(\os)$, $N(\mgx)/N(\os)$ and $N(\mgx)/N(\neo)$, within their 10\% 
uncertainty, are well reproduced for $\log$~U $\sim$ 1.0 and 1.1 respectively. However, 
ratios involving low ions (i.e. \ofo, \nfo, \of) require lower values of $\log$~U 
indicating the presence of multiple gas phases. By using $f_c = 0.6$, the covering 
fraction determined for \neo, we obtain an upper limit of log~$N(\hi)$ = 14.36 for 
the \hi\ column density associated with the high ionization gas.This gives a lower 
limit on metallicity of log $(Z/Z_{\odot}) \gtrsim -0.8$ for component 3. For all 
the components, the observed range of $N(\neo)/N(\os)$ ratio is consistent with 
0.95 $\le$ log~U $\le$ 1.30.

Because of poor $S/N$ of the $FUSE$ data and the Galactic H$_2$ contamination we 
could only check the consistency of \mgx\ profiles predicted by the models for 
other components. For each component we calculate the model predicted $N(\mgx)$ 
for log~U required to produce the observed $N(\neo)/N(\os)$. The profiles are then 
generated assuming $b$ and $f_c$ as those measured for \neo. The predicted \mgx\ 
profile is consistent with the observed spectra within measurement uncertainties 
(see Fig.~\ref{vplot}) suggesting similar value of log~U for all these 
components spread over $\sim$1800 \kms. 

Nitrogen is reported to be over abundant compared to oxygen in associated absorbers 
favoring a rapid enrichment scenario in the central regions of QSOs 
\citep[][]{Hamann92,Korista96,Petitjean99}. The best fit photoionization model 
predicts log~$N(\nf) \sim 13.5$. Even in a low dispersion spectrum such a line 
should be detectable at $\sim 4 \sigma$ level. The fact that we do not detect 
\nf\ line could be because (a) N is not over abundant compared to O and/or (b) 
the covering fraction of \nf\ may be much less as it falls on top of Ly$\alpha$+\nf\ 
emission lines. High resolution data is needed to confirm the above mentioned 
possibilities.

\begin{table}
\caption{Partial coverage corrected Voigt profile fit parameters.} 
\begin{tabular}{cccccc} 
\hline 
Species & $v_{\rm rel}$(\kms)$^{~1}$ & ID$^{~2}$ & $b$(\kms) & log~$N$(cm$^{-2}$) & $f_c^{~3}$\\  
\hline 
\neo  & $-$5767 $\pm$ 9   & (1) & 143.8 $\pm$ 12.7  & 14.22 $\pm$ 0.03 & 1.0 \\
\hi   &        .....      &     & 143.8             & $\le$13.71 & 1.0\\ 
\hline 
\os   & $-$4743 $\pm$ 20  & (2) & 127.1 $\pm$ 20.2  & 14.67 $\pm$ 0.09& 0.5 \\ 
\neo  &          .....    &     & 127.1 $\pm$  0.0  & 15.14 $\pm$ 0.02& 0.6 \\
\hi   &          .....    &     & 127.1             & $\le$ 13.70& 1.0 \\ 
\hline 
\os\  & $-$4539  $\pm$ 6  & (3) &  98.4 $\pm$  5.7  & 15.11 $\pm$ 0.03& 0.5 \\
\neo\ &          .....    &     &  96.7 $\pm$  1.9  & 15.45 $\pm$ 0.01& 0.6 \\
\ofo  &          .....    &     &  98.4             & $\le$ 13.44     & 1.0 \\
\of   &          .....    &     &  98.4             & 14.45           & 0.5 \\ 
\nt   &          .....    &     &  98.4             & $\le$ 14.29     & 0.5 \\ 
\nfo  &          .....    &     &  98.4             & 13.61 $\pm$ 0.04& 0.5 \\ 
\mgx  &          .....    &     &  98.4             & 15.32           & 0.6 \\ 
\hi   &          .....    &     &  98.4             & 14.36(15.75)    & 0.6(0.1) \\ 
\hline 
\neo\ & $-$3678 $\pm$ 4   & (4) & 94.9 $\pm$ 3.5 & 15.10 $\pm$ 0.02 & 0.6\\ 
\os\  &          .....    &     & 96.8 $\pm$ 3.9 & 14.84 $\pm$ 0.02 & 0.6\\
\ofo  &          .....    &     & 96.8           & 14.39            & 0.6\\ 
\nt   &          .....    &     & 94.9           & 13.79 $\pm$ 0.13 & 0.6\\
\nfo  &          .....    &     & 94.9           & 13.55 $\pm$ 0.04 & 0.6\\ 
\hi   &          .....    &     & 94.9           & $\le$ 14.11      & 1.0\\ 
\hline 
\neo  & $-$3491 $\pm$ 2   & (5) & 63.8 $\pm$ 2.9 & 15.11 $\pm$ 0.02 & 0.6\\
\os   &          .....    &     & 68.3 $\pm$ 5.4 & 14.55 $\pm$ 0.03 & 0.6\\ 
\of   &          .....    &     & 68.3           & 14.03            & 0.6\\ 
\mgx  &          .....    &     & 68.3           & 15.11            & 0.6\\
\hi   &          .....    &     & 63.8           & $\le$ 14.02      & 1.0\\ 
\hline 
\neo\ & $-$3354 $\pm$ 4   & (6) & 37.8 $\pm$ 4.5 & 14.70 $\pm$ 0.06 & 0.6\\
\os\  &          .....    &     & 39.0 $\pm$ 5.7 & 14.05 $\pm$ 0.06 & 0.6\\ 
\hi   &          .....    &     & 37.8           & $\le$ 13.91      & 1.0\\ 
\hline 
\neo\ & $-$3285 $\pm$ 15  & (7) & 40.6 $\pm$13.9 & 14.13 $\pm$ 0.21 & 0.6\\ 
\hi   &          .....    &     & 40.6           & $\le$ 13.94      & 1.0\\ 
\hline
\end{tabular}
~~~~~~~~~~~~~~~~~~~~~~~~~~~~~~~~~~~~~~~~~~~~~~~~~~~~~~~~~~~~~~~~~~~~~~~~~~~
~~~~~~~~~~~~~~~~~~~~~~~~~~~~~~~~~~~~~~~~~~~~~~~~~~~~~~~~~~~~~~~~~~~~~~~~~~~
Notes -- $^{1}$ Relative velocity with respect to \zem. $^{2}$ Component ID as in 
Fig.~\ref{vplot}. $^{3}$ covering fraction used for the fit. 
Rest frame wavelengths and oscillator strengths are taken from \citet{Verner94}. 
\label{tab_cl}  
\end{table}  

\section{Summary \& Conclusions}  
\label{con}

We report the detection of highly ionized outflowing gas through 
associated absorption from \os, \neo\ and \mgx\ in the UV spectra of 
HE~0238$-$1904 which is presumably a non-BALQSO.  
The high $S/N$ COS spectrum has allowed us to determine the covering fraction 
($f_c$) and the multi-phase physical conditions in the absorber. The similarity 
in the absorption profiles and $f_c$ suggest that \os\ and \neo\ absorption 
possibly trace the same phase of the absorbing gas. Near constancy of $f_c$ for 
\neo\ and \os\ in the different components spread over $\sim$1800 \kms\ indicates  
that the flow could be collimated. 

\begin{figure} 
\centerline{
\vbox{
\includegraphics[height=7.0cm,width=9.0cm,angle=00]{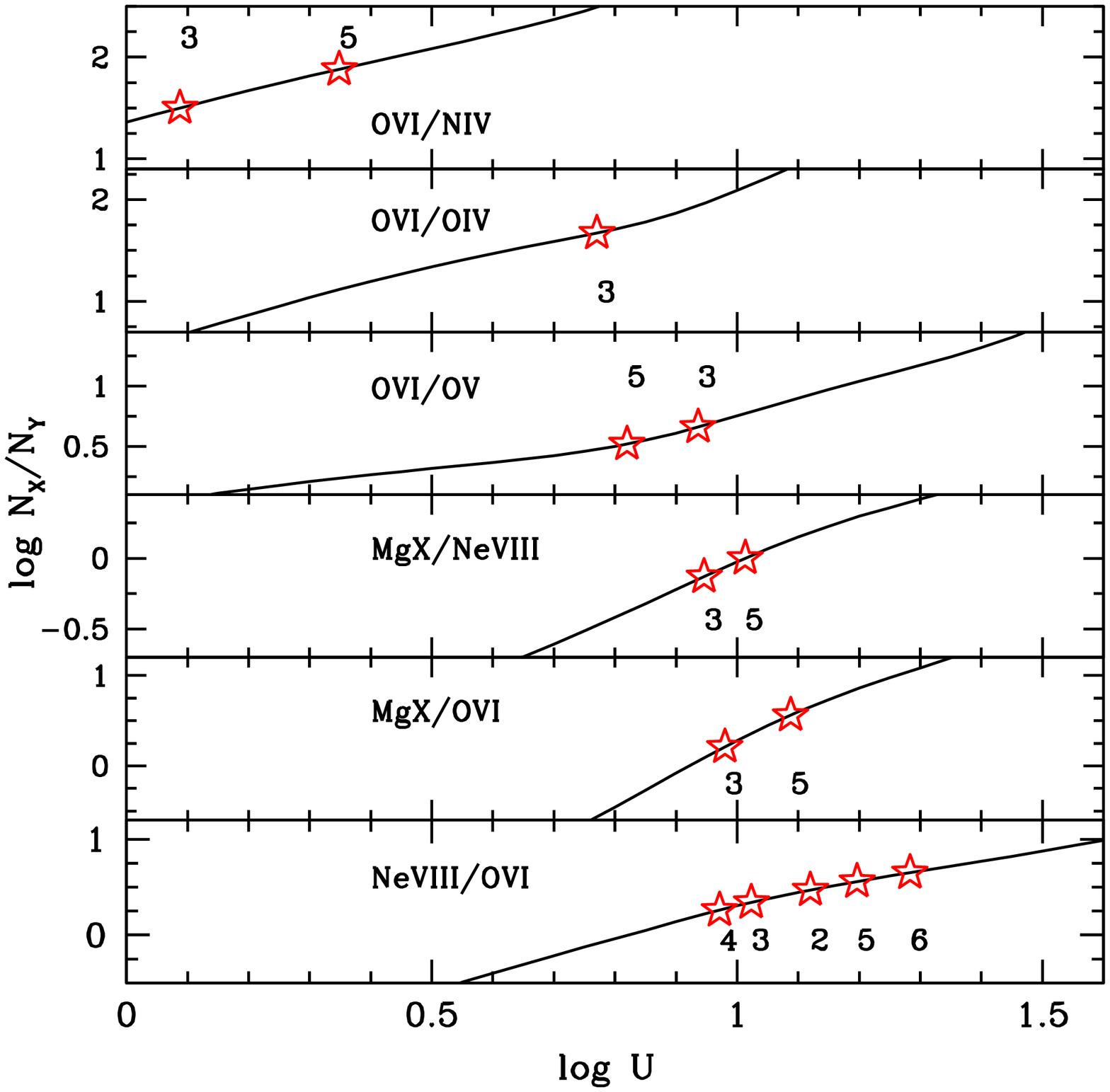} 
\centerline{\hbox{ 
\includegraphics[height=4.0cm,width=8.8cm,angle=00]{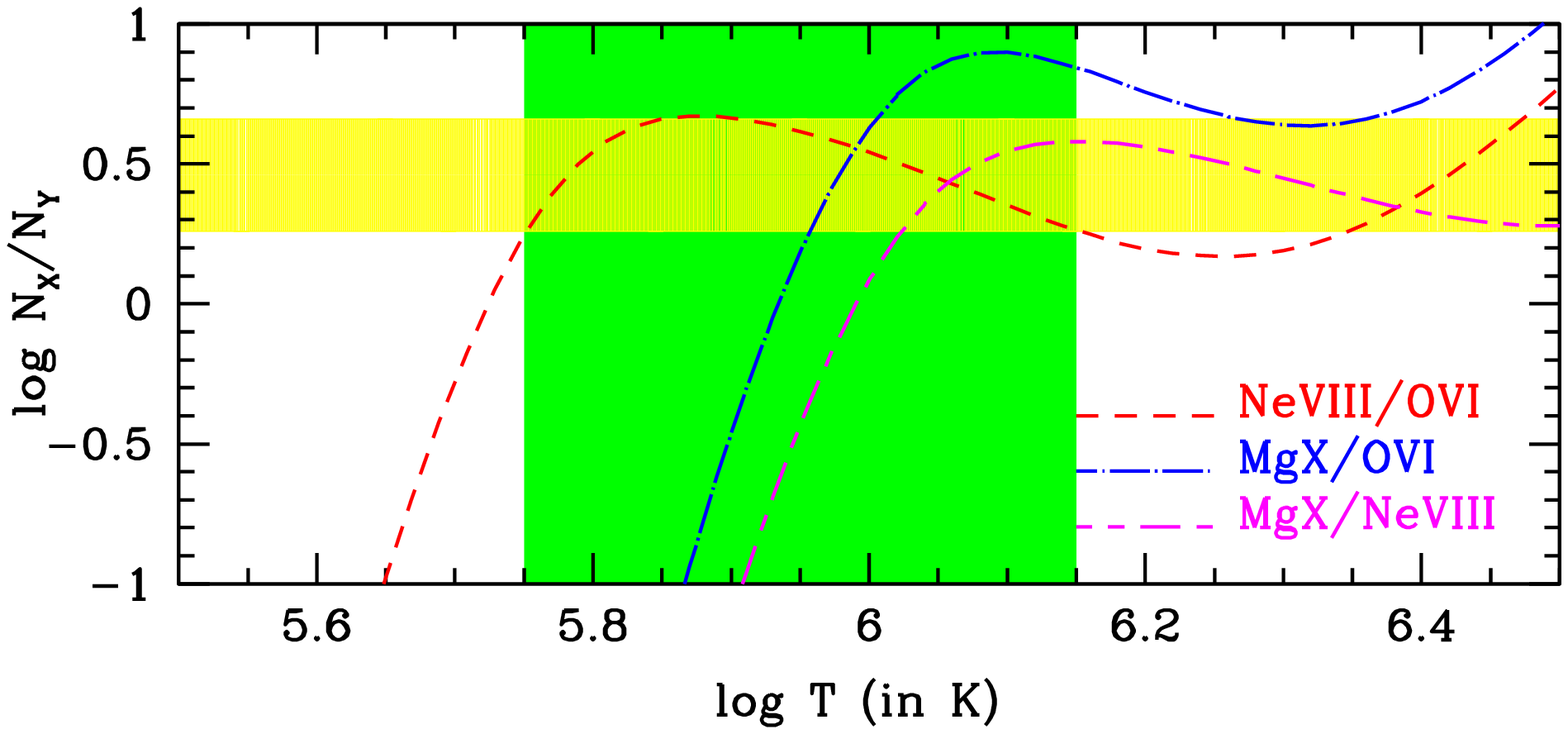} 
}}
}}
\caption{{Top:} Results of photoionization calculation. In each sub-panel the 
solid curve shows the Cloudy predicted column density ratios against log U. 
In the bottom most panel the measurements of $N(\neo)/N(\os)$ ratios in five 
components are shown by the stars together with their component ID. 
For other panels ratios are shown only for component 3 \& 5.  
{Bottom:} Column density ratios versus temperature under collisional ionization 
equilibrium \citep[]{Gnat07}. 
The horizontal shaded region shows the range in the measured 
$N(\neo)/N(\os)$ whereas the vertical shaded region shows corresponding allowed 
range in temperature. Clearly within the permissible range of temperature, 
$N(\mgx)/N(\os)$ and/or $N(\mgx)/N(\neo)$ ratios can vary more than an order 
of magnitude.   
} 
\label{model} 
\end{figure}

Under photoionization equilibrium conditions, the log~$N(\neo)/N(\os)$ = 0.46$\pm$0.16 
ratio in the different components is consistent with 0.95 $\le$ log~U $\le$ 1.30. 
This means that in spite of the absorbing gas being distributed over $\sim$1800 \kms, 
the product n$_{\rm H}$r$_{\rm c}^{2}$ (where r$_{\rm c}$ is the distance 
of the gas from the QSO) is nearly a constant. Getting such constraint naturally for 
different components could be an issue for the photoionization model. Models of mass 
conserving shells expanding with a velocity $v_s$ predict 
n$_{\rm H}$r$_{\rm c}^{2}$$v_s$ to be a constant. However, this is not the 
case in the present system as we do not find any clear trend between the ejection 
velocity and the ionization parameter. Alternatively, collisional ionization in gas 
with temperatures in the range $5.7 \lesssim {\rm log}~T (\rm K) \lesssim 6.1$ can 
recover the 
observed $N(\neo)/N(\os)$ ratio as shown in the bottom panel of Fig~\ref{model}. 
In this case we expect $N(\mgx)/N(\neo)$ or $N(\mgx)/N(\os)$ to be very different 
between the different components, which cannot be ruled out with the existing 
$FUSE$ data on \mgx. We note here that even in the case of photoionization the 
equilibrium gas temperature is $\sim 1.2\times10^{5}$~K for log~U $\sim$ 1.0.   

From the observed flux at the Lyman limit 
(1.4$\times 10^{-14}$~erg cm$^{-2}$ s$^{-1}$ \AA$^{-1}$) in the QSO rest frame we 
get a relation log~(n$_{\rm H}$r$_{\rm c}^{2}$) = 45.8 $-$ log~U. This 
gives log~(n$_{\rm H}$r$_{\rm c}^{2}$) = 44.8 for log~U = 1.0. If we 
assume that the absorbing gas is well within the BLR (r$_{\rm c} \leq R_{BLR}$) 
we get n$_{\rm H} > 5 \times 10^8$~cm$^{-3}$. The cloud thickness along the 
line of sight is, $\Delta \rm r_{\rm cl} \leq 2 \times 10^{12}$ cm, if we use 
$N(\hi) \sim 10^{14}$ cm$^{-2}$ and neutral fraction of $f_{\hi} \sim 10^{-7}$ as 
found for the photoionization model at log~U = 1.0. The fact that the cloud 
is covering $\sim$ 60\% of the background emitting source gives the transverse size 
of the cloud as 
r$_{\rm cl} \sim D_{2700} \sqrt{f_c}/2$  = 1.2$\times$10$^{17}$ cm$^{-2}$. 
This is $\sim$ 5 orders of magnitudes larger than the thickness along the line of 
sight (i.e. $\Delta \rm r_{\rm cl}/r_{\rm cl} \sim 10^{-5}$) resembling  a 
sheet like geometry.  

If we assume spherical geometry for the absorbing gas then the observed $f_c$ 
suggests that the radius of the cloud is $\sim 1.2 \times 10^{17}$ cm. 
This gives n$_{\rm H} \sim 10^4$ cm$^{-3}$ and r$_{\rm c} \sim 90$ pc for 
log~U = 1.0. Therefore the absorbing region, if nearly 
spherical, will be co-spatial with the narrow emission line region (NLR) gas. The sound 
crossing time is $\sim 4$ days and $\sim 700$ years respectively, for the two 
scenarios discussed above. Purely from stability point of view, the absorbing gas 
is likely to be outside the BLR \citep[see][]{Faucher11}.The NRAO VLA Sky Survey 
(NVSS) image shows possible radio emission associated with the QSO. We speculate 
that the outflow triggered by the radio jet could possibly explain the collimated 
hot outflowing gas outside the BLR as seen in the case of 3C48 \citep[]{Gupta05}. 

The gas phase metallicity is found to be, Z $\ge$ 0.2 Z$_\odot$ without a clear 
evidence for an enhanced N abundance. However, high resolution spectra covering 
\cf\ and \nf\ is needed  to perform detailed abundance analysis. High densities 
(n$_{\rm H}$) inferred for the system suggests that HE~0238$-$1904 is an ideal 
target for absorption line variability studies. However, we do not find any 
signature of strong variation in \neo\ and \os\ equivalent width from 2002 
onwards from the $HST$/STIS and COS spectra. As this source is X-ray  bright 
detection of warm absorbers will allow us to get further insight into this hot 
outflowing gas. 

\section{acknowledgment} 
We appreciate the efforts of the many people involved with the design and 
construction of COS \& FUSE and its deployment on the HST. Thanks are also 
extended to the people responsible for determining the orbital performance 
of COS and developing the CalCOS data processing pipeline. We wish to 
acknowledge the IUCAA/IGO staff for their support during our observations. 
We thank Pushpa Khare for careful reading of the manuscript. SM thanks CSIR 
for providing support for this work.  

\def\aj{AJ}%
\def\actaa{Acta Astron.}%
\def\araa{ARA\&A}%
\def\apj{ApJ}%
\def\apjl{ApJ}%
\def\apjs{ApJS}%
\def\ao{Appl.~Opt.}%
\def\apss{Ap\&SS}%
\def\aap{A\&A}%
\def\aapr{A\&A~Rev.}%
\def\aaps{A\&AS}%
\def\azh{AZh}%
\def\baas{BAAS}%
\def\bac{Bull. astr. Inst. Czechosl.}%
\def\caa{Chinese Astron. Astrophys.}%
\def\cjaa{Chinese J. Astron. Astrophys.}%
\def\icarus{Icarus}%
\def\jcap{J. Cosmology Astropart. Phys.}%
\def\jrasc{JRASC}%
\def\mnras{MNRAS}%
\def\memras{MmRAS}%
\def\na{New A}%
\def\nar{New A Rev.}%
\def\pasa{PASA}%
\def\pra{Phys.~Rev.~A}%
\def\prb{Phys.~Rev.~B}%
\def\prc{Phys.~Rev.~C}%
\def\prd{Phys.~Rev.~D}%
\def\pre{Phys.~Rev.~E}%
\def\prl{Phys.~Rev.~Lett.}%
\def\pasp{PASP}%
\def\pasj{PASJ}%
\def\qjras{QJRAS}%
\def\rmxaa{Rev. Mexicana Astron. Astrofis.}%
\def\skytel{S\&T}%
\def\solphys{Sol.~Phys.}%
\def\sovast{Soviet~Ast.}%
\def\ssr{Space~Sci.~Rev.}%
\def\zap{ZAp}%
\def\nat{Nature}%
\def\iaucirc{IAU~Circ.}%
\def\aplett{Astrophys.~Lett.}%
\def\apspr{Astrophys.~Space~Phys.~Res.}%
\def\bain{Bull.~Astron.~Inst.~Netherlands}%
\def\fcp{Fund.~Cosmic~Phys.}%
\def\gca{Geochim.~Cosmochim.~Acta}%
\def\grl{Geophys.~Res.~Lett.}%
\def\jcp{J.~Chem.~Phys.}%
\def\jgr{J.~Geophys.~Res.}%
\def\jqsrt{J.~Quant.~Spec.~Radiat.~Transf.}%
\def\memsai{Mem.~Soc.~Astron.~Italiana}%
\def\nphysa{Nucl.~Phys.~A}%
\def\physrep{Phys.~Rep.}%
\def\physscr{Phys.~Scr}%
\def\planss{Planet.~Space~Sci.}%
\def\procspie{Proc.~SPIE}%
\let\astap=\aap
\let\apjlett=\apjl
\let\apjsupp=\apjs
\let\applopt=\ao
\bibliographystyle{mn}
\bibliography{mybib}
\end{document}